\documentclass{PoS}

\usepackage[T1]{fontenc}
\usepackage{xspace}
\usepackage{floatrow}
\newcommand{\gambit}{GAMBIT\xspace}

\newcommand{\GB}{\gambit}

\title{Dark matter theory: Implications and future prospects for Fermi}

\ShortTitle{Dark matter theory: Implications and future prospects for Fermi}

\author{\speaker{Pat Scott}\\
Fundamental Physics Section, Department of Physics, Imperial College London, Blackett Laboratory, Prince Consort Road, London SW7 2AZ, UK\\
E-mail: \email{p.scott@imperial.ac.uk}}

\abstract{I give a brief review of some of the implications of Fermi data for theories of the identity of dark matter, and their combination with data from other complementary probes. I also preview some of the prospects for probing such models with future data.}

\FullConference{7th Fermi Symposium 2017\\
		15-20 October 2017\\
		Garmisch-Partenkirchen, Germany}

\begin{document}

\section{Introduction}

Even the simplest theories for dark matter (DM) invariably have a wide range of theoretical, experimental and computational implications.  Their interplay very often means that a single search for DM or new physics cannot actually be meaningfully interpreted in isolation from complementary searches or detailed theoretical calculations.

To understand the implications of $\gamma$-ray data for any particular theory of DM, it is important to consider a number of inherently model-dependent aspects.  These include the specific annihilation branching fractions of the DM candidate, the temperature-dependence of its annihilation cross-section and corresponding thermal relic abundance, and complementary constraints from direct detection, collider searches, flavour physics and other indirect searches for DM.

\subsection{An example: scalar singlet annihilation to $\gamma$-ray lines}

A concrete example of the need to carefully and consistently consider both theoretical and complementary experimental implications of the model at hand can be seen in the scalar singlet model for DM \cite{SilveiraZee,McDonald94,Burgess01,Cline13b,SSDM}, and its loop-induced annihilation to two photons.  This model adds one real scalar beyond the Standard Model (BSM), uncharged under any of the SM gauge groups, but stabilised by the imposition of an additional $\mathbb{Z}_2$ symmetry.  The singlet $S$ interacts with the SM Higgs, and indirectly, with the rest of the SM, via the so-called `Higgs portal' coupling $\lambda_{hS} S^2H^2$.  This interaction leads to nuclear scattering on nucleons (direct detection), self-annihilation to SM final states (indirect detection) and thermal production in the early Universe.  It also induces the Higgs to decay to two $S$ bosons whenever kinematically possible.

A number of authors \cite{2010PhRvD..82l3514P,2015JHEP...03..045F,2015PhLB..751..119D} have computed the relevant matrix elements and corresponding line signals at \textit{Fermi}-LAT for the loop-induced process $SS\to \gamma\gamma$, claiming that it provides significant constraints on the model.  This monochromatic final state is appealing because it is easily discriminated from astrophysical backgrounds, leading to strong limits \cite{LATLineP8}.  The cross-section is (unsurprisingly) maximised at large $\lambda_{hS}$.  Because the annihilation proceeds via an intermediate SM Higgs, it is also maximised where the $S$ annihilates on resonance with the Higgs, at $m_S \sim m_h/2$.  However, annihilation to all other final states is also increased in exactly the same areas of parameter space, causing the thermal relic abundance to lie well below the observed cosmological abundance of DM in all cases where the line signal is potentially observable.  Accounting for the small fraction of DM in $S$ bosons shows that in fact, the line signal is never observable in this model, and does not provide any constraint on it.

Non-thermal production of $S$ would allow it to be all of DM even when the thermal abundance is depleted -- but this requires the introduction of additional heavy states not included in the original model.  A similar criticism can be levelled at the assumption that the \textit{rest} of DM is made up of something other than the $S$ boson. This serves to illustrate the fact that the exact definition of the model under consideration can completely change any conclusions that one draws from $\gamma$-ray data.

\subsection{Progress interpreting indirect searches}

When trying to understand the implications for a given DM model of a single search, treating all relevant theoretical and phenomenological aspects of the model consistently is a rather non-trivial task.  Experimental results in indirect detection need to be recast from published limits on toy models with 100\% annihilation into a single final state, to implied limits on the specific combination of branching fractions exhibited by real models in each part of their parameter spaces (see e.g.\ \cite{Scott09c,IC22Methods,IC79_SUSY}).  Self-consistently accounting for the implications of other experiments on the same model requires a similar recast of those other results, whether they be direct, indirect or collider searches.  Accounting for theoretical aspects, such as the changing thermal relic density across a model's parameter space, requires detailed computations, which must be repeated for each combination of parameter values.  Ideally, one would also account for uncertainties on the various input parameters to these calculations (such as SM and nuclear parameters, DM halo models, etc), both theoretical and experimental systematics, and numerical challenges arising from the difficulty of sampling such complicated parameter spaces (see e.g.\ Refs.\ \cite{Akrami09,SBcoverage,Akrami11Coverage,SBSpike,Strege12}).  Repeating this exercise for many different DM models, and comparing the results in a systematic way, becomes a phenomenological exercise of mammoth proportions.

This problem is addressed by GAMBIT (The Global and Modular BSM Inference Tool) \cite{gambit}, a flexible and adaptive framework for computing the combined implications of all theoretical aspects and experimental searches sensitive to a given model for DM or new physics.  It includes detailed observable and likelihood libraries for DM searches \cite{DarkBit}, collider \cite{ColliderBit} and flavour \cite{FlavBit} experiments, precision, decay and mass-spectrum data \cite{SDPBit}, as well as a dedicated statistical analysis module \cite{ScannerBit}.  The DM module features searches for annihilation in dwarf galaxies by \textit{Fermi} \cite{LATdwarfP8}, and in the direction of the Galactic Centre with various experiments \cite{Abramowski:2011hc,Calore14,Silverwood14}.

\section{Implications of \textit{Fermi} and other searches for scalar singlet DM}

\begin{figure}
\centering
\includegraphics[trim=20 410 430 40, clip=true, width=\textwidth]{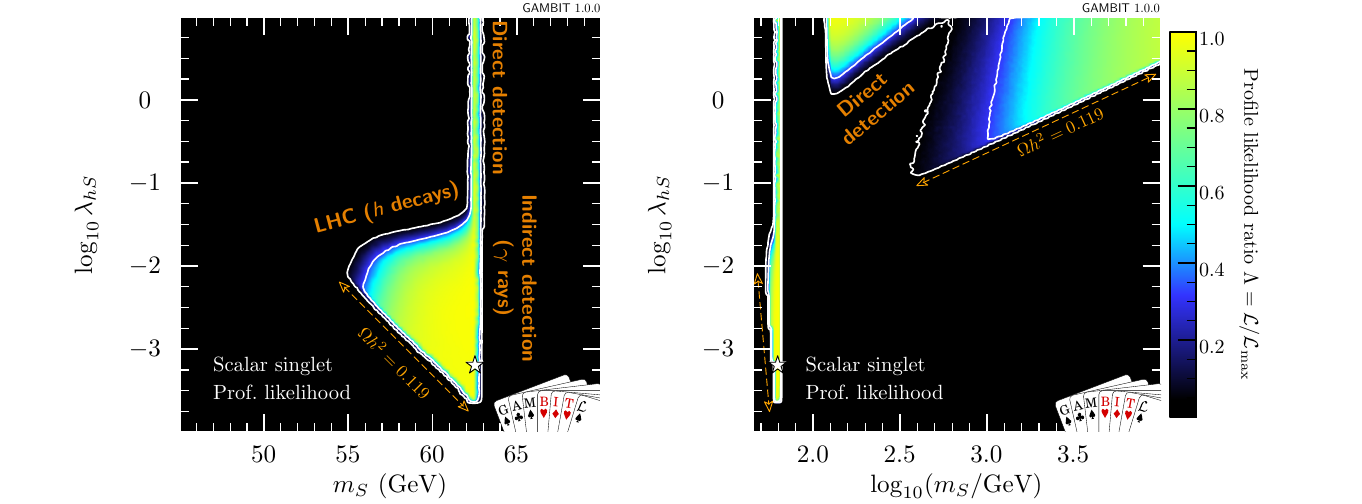}
\caption{Allowed parameters of scalar singlet DM \cite{SSDM}, in terms of its mass $m_S$ and coupling to the Higgs $\lambda_{hS}$.  The left panel shows a zoomed-in view of the low-mass region of the right panel.  White contours indicate 68\% and 95\% confidence profile likelihood regions, and shading indicates the value of the normalised profile likelihood.  All direct and indirect signals have been rescaled to self-consistently account for the thermal relic abundance of scalar singlet particles.  Orange labels at the edges of the allowed regions indicate the leading experimental constraint applying in that area of the parameter space, as well as the specific edges where the singlet accounts for the entirety of the observed abundance of DM.  White stars locate the best fit.}
\label{singlet}
\end{figure}

Fig.\ \ref{singlet} shows the global implications for the scalar singlet DM model of all relevant experimental searches and theoretical constraints \cite{SSDM}.  Preferred regions are plotted in terms of the two BSM parameters of the theory: the portal coupling $\lambda_{hS}$ and DM mass $m_S$, with the left panel providing a zoomed-in view of the low-mass region of the right panel.  Dominant constraints are indicated by orange text to each side of the preferred regions.  Two regions persist after all constraints have been applied: a low-mass `resonant triangle' associated with on-resonance annihilation of the $S$, leading to a suppression of the relic density (and therefore also of direct and indirect signals), and a high-mass region beyond the current reach of direct detection.

\floatsetup[figure]{capposition=beside,capbesideposition={center,right}}
\begin{figure}
\includegraphics[width=0.49\textwidth]{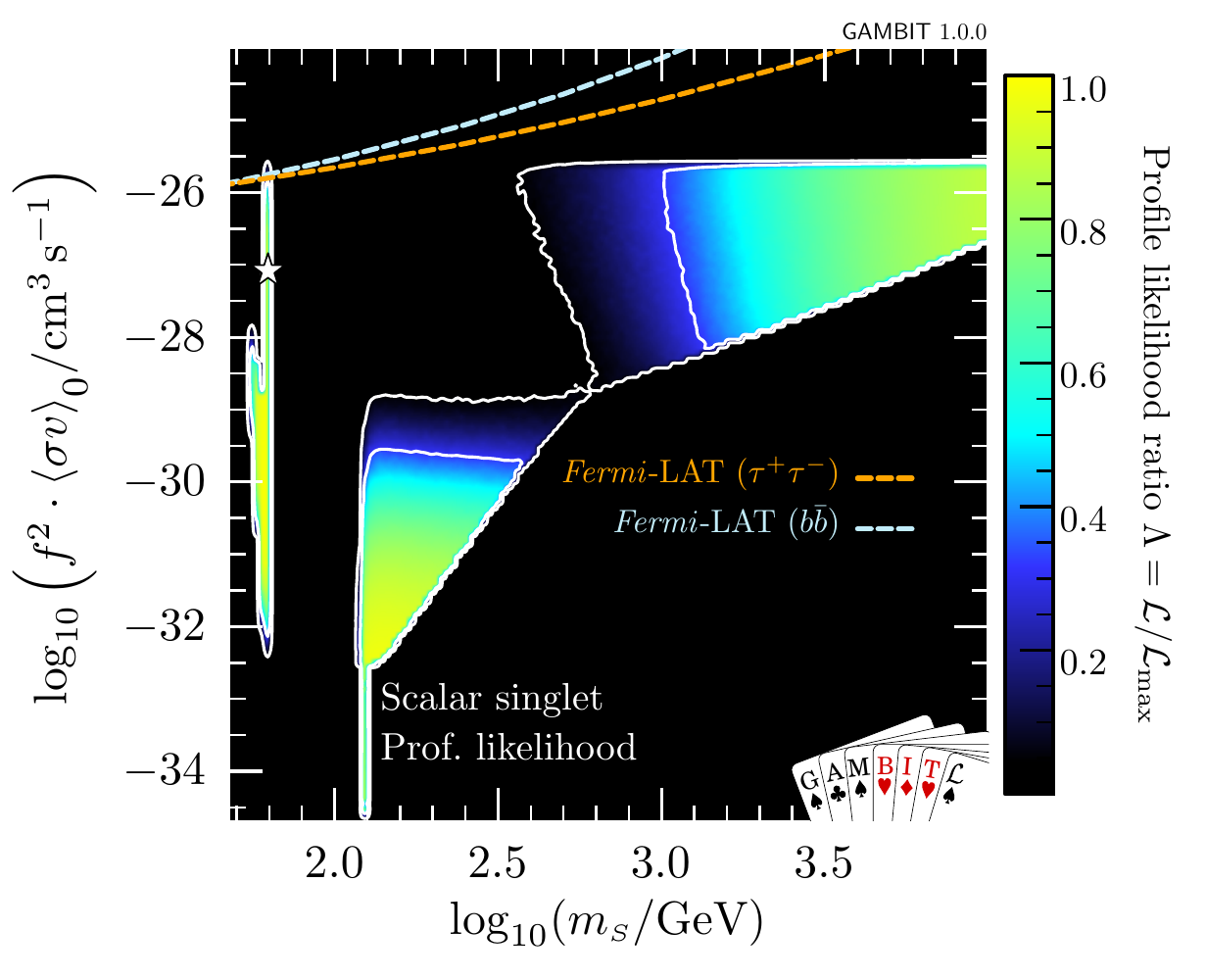}
\caption{As per Fig.\ \protect\ref{singlet}, but expressed in terms of the rescaled annihilation cross-section.  For display purposes the annihilation cross-section has been rescaled in compensation for the scaling of direct and indirect signals for the relic density, to allow direct comparison with the official limits from \textit{Fermi}-LAT searches for DM annihilation in dwarf spheroidal galaxies \cite{LATdwarfP8}.}
\label{singlet_ann}
\end{figure}
\floatsetup[figure]{capposition=below}

For comparison, the same results are shown in terms of the total annihilation cross-section in Fig.\ \ref{singlet_ann}.  Note that models that produce less $S$ particles during thermal freezeout than the total observed abundance of DM are not penalised.  As the BSM parameter space is only two-dimensional, the interior of the preferred regions have $S$ as a sub-dominant component of DM.  At some edges, $S$ constitutes all of DM; these are explicitly indicated in Fig.\ \ref{singlet}.  Where $S$ is not all of DM, the predicted signals at direct and indirect detection experiments have been rescaled to account for the reduced $S$ abundance.  To factor out this suppression (in order to make an illustrative comparison with the \textit{Fermi} limits included in the global analysis), the cross-sections plotted Fig.\ \ref{singlet_ann} are rescaled by the square of the fraction $f=\Omega_S/\Omega_{\rm DM}$ of the relic abundance of DM constituted by the singlet.\footnote{The square in this factor is due to the quadratic dependence of the $\gamma$-ray flux on the density of DM.}  We see here that limits from \textit{Fermi} searches for annihilation in dwarf galaxies (dashed lines; \cite{LATdwarfP8}) constrain only the upper edge of the resonance with the SM Higgs, where thermal effects suppress the annihilation cross-section in the early Universe, but not locally, where the \textit{Fermi} searches apply.

\begin{figure}
\includegraphics[width=0.49\textwidth]{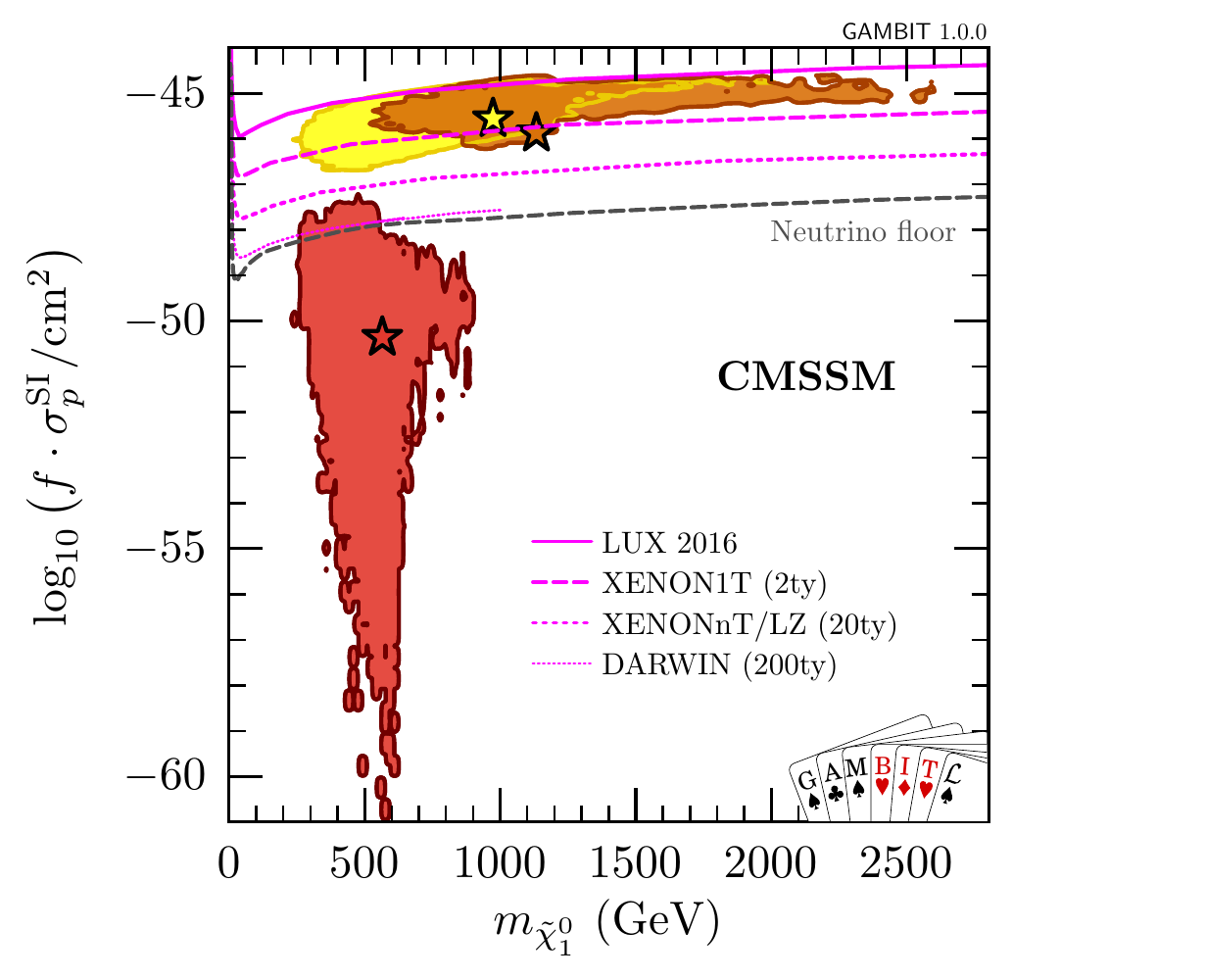}
\includegraphics[width=0.49\textwidth]{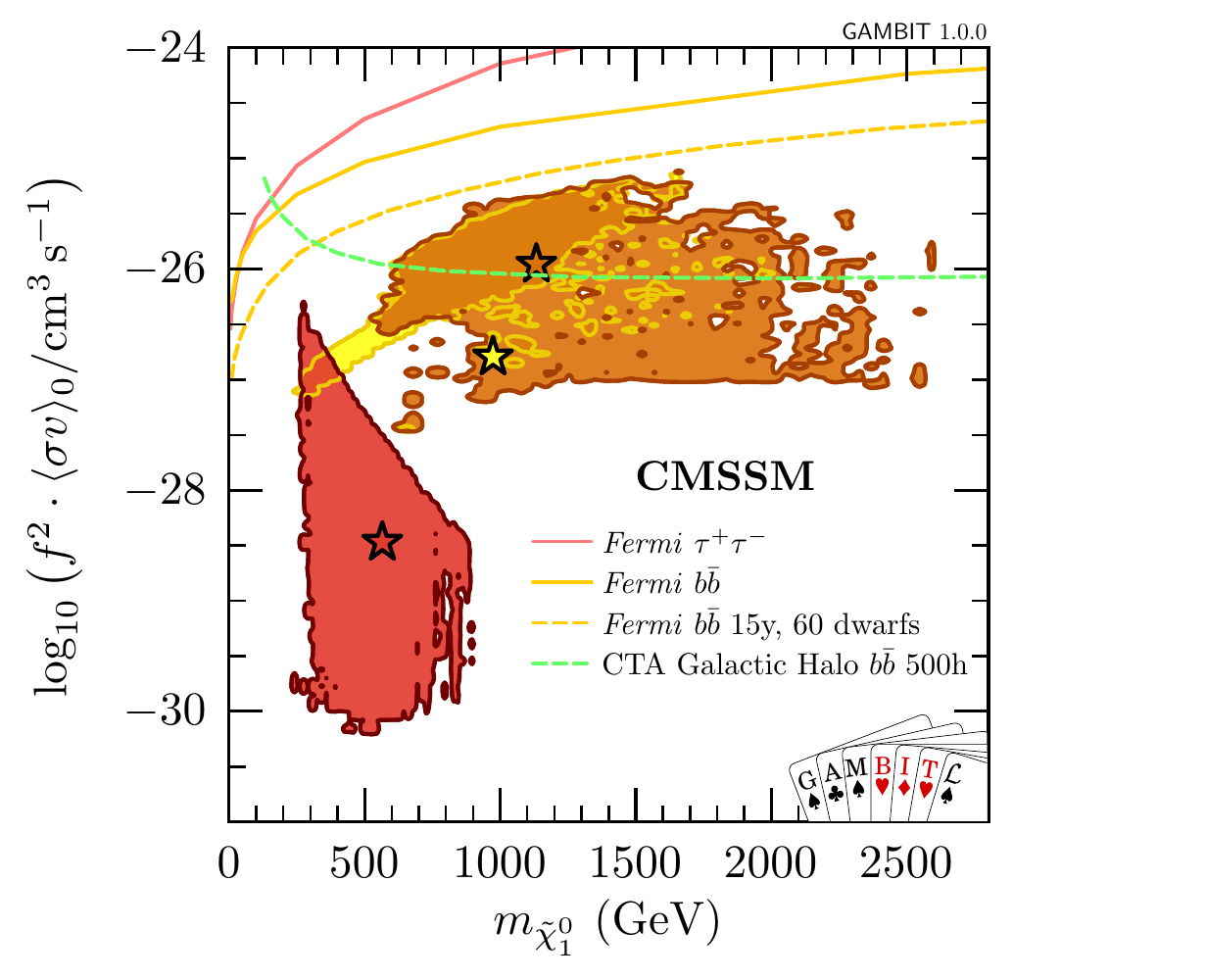}\\
\includegraphics[height=0.3cm]{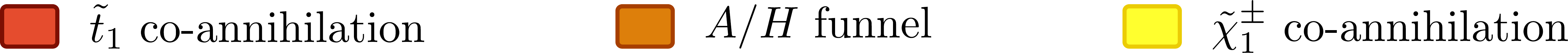}
\caption{95\% confidence regions of the Constrained Minimal Supersymmetric SM (CMSSM) \cite{CMSSM}, in terms of the DM mass $m_{\chi^0_1}$ and its rescaled spin-independent scattering cross-section with protons (\textit{left}) or self-annihilation cross-section (\textit{right}).  Shading indicates processes able to deplete the relic density. Stars are the best fits in each region.  Direct and indirect signals have been rescaled to self-consistently account for the thermal DM relic abundance, but for display, the cross-sections have been rescaled in compensation, to allow comparison with official 90\% confidence limits from LUX \cite{LUXrun2}, 95\% limits from \textit{Fermi}-LAT searches towards dwarf galaxies \cite{LATdwarfP8}, and projections from XENON \cite{XENONnTLZ}, DARWIN \cite{DARWIN}, \textit{Fermi} \cite{Charles:2016pgz} and CTA \cite{Carr:2015hta}.  The dashed grey line indicates the expected `neutrino floor' for direct detection \cite{Billard:2013qya}.
}
\label{CMSSM}
\end{figure}

\begin{figure}
\includegraphics[width=0.49\textwidth]{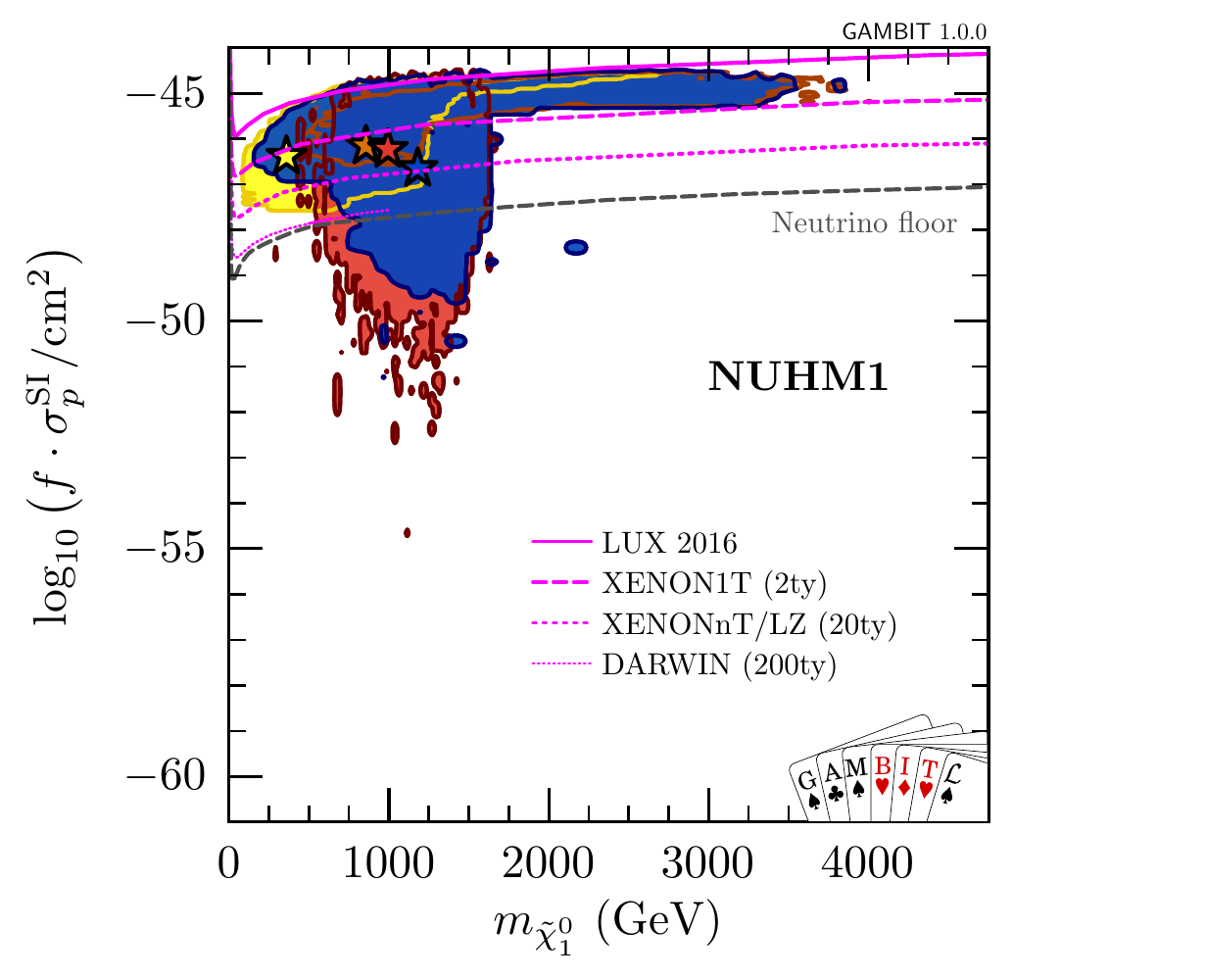}
\includegraphics[width=0.49\textwidth]{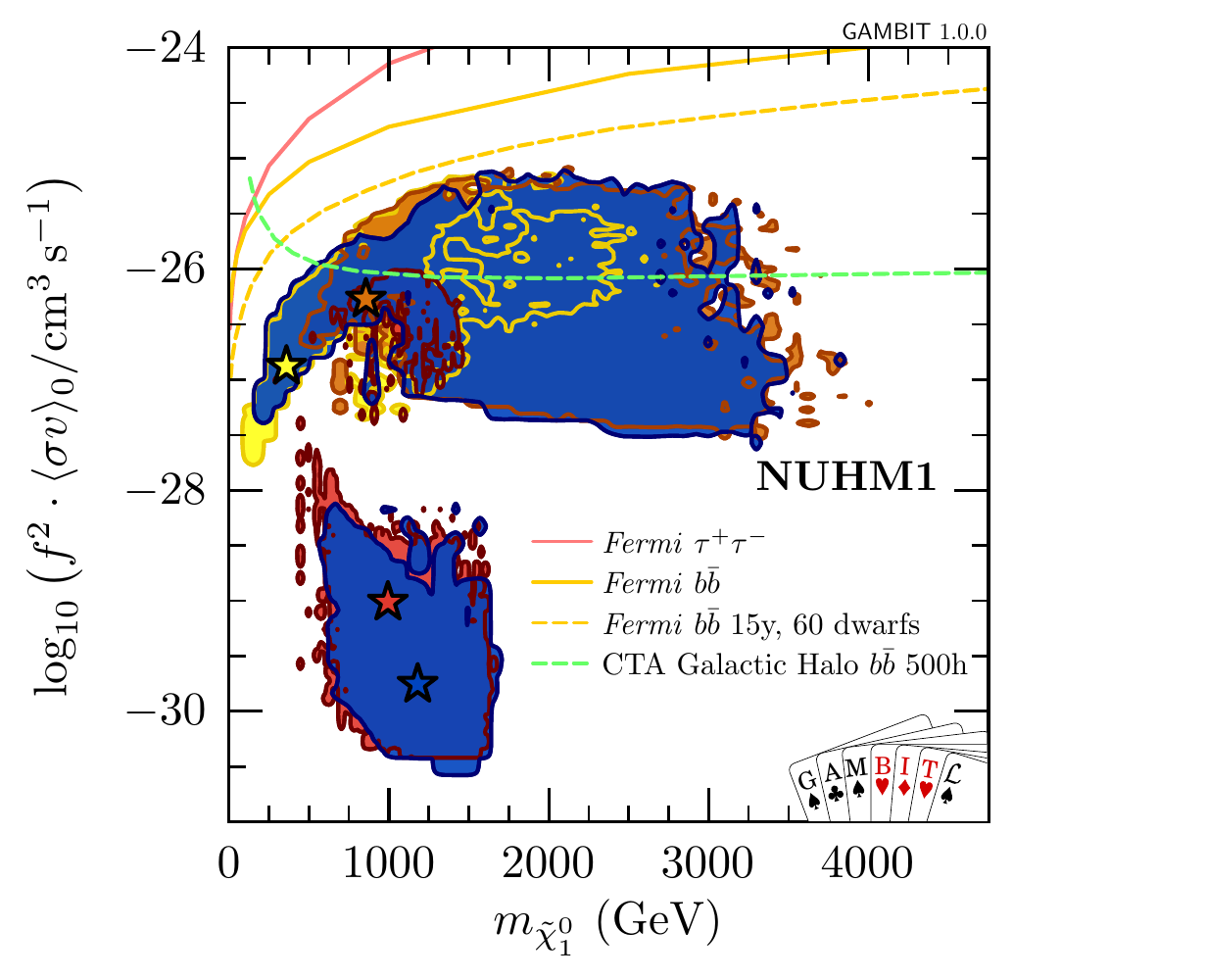}\\
\includegraphics[width=0.49\textwidth]{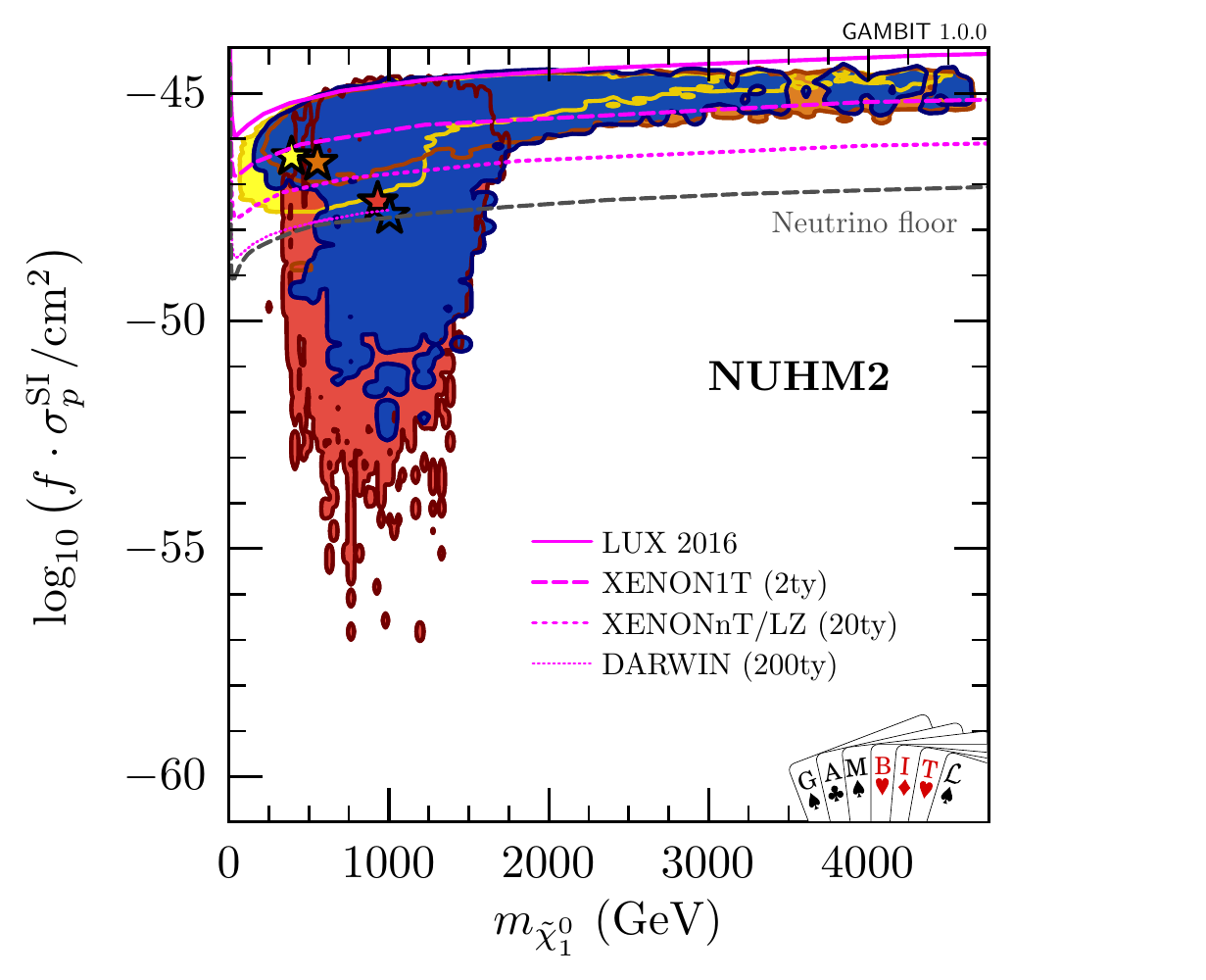}
\includegraphics[width=0.49\textwidth]{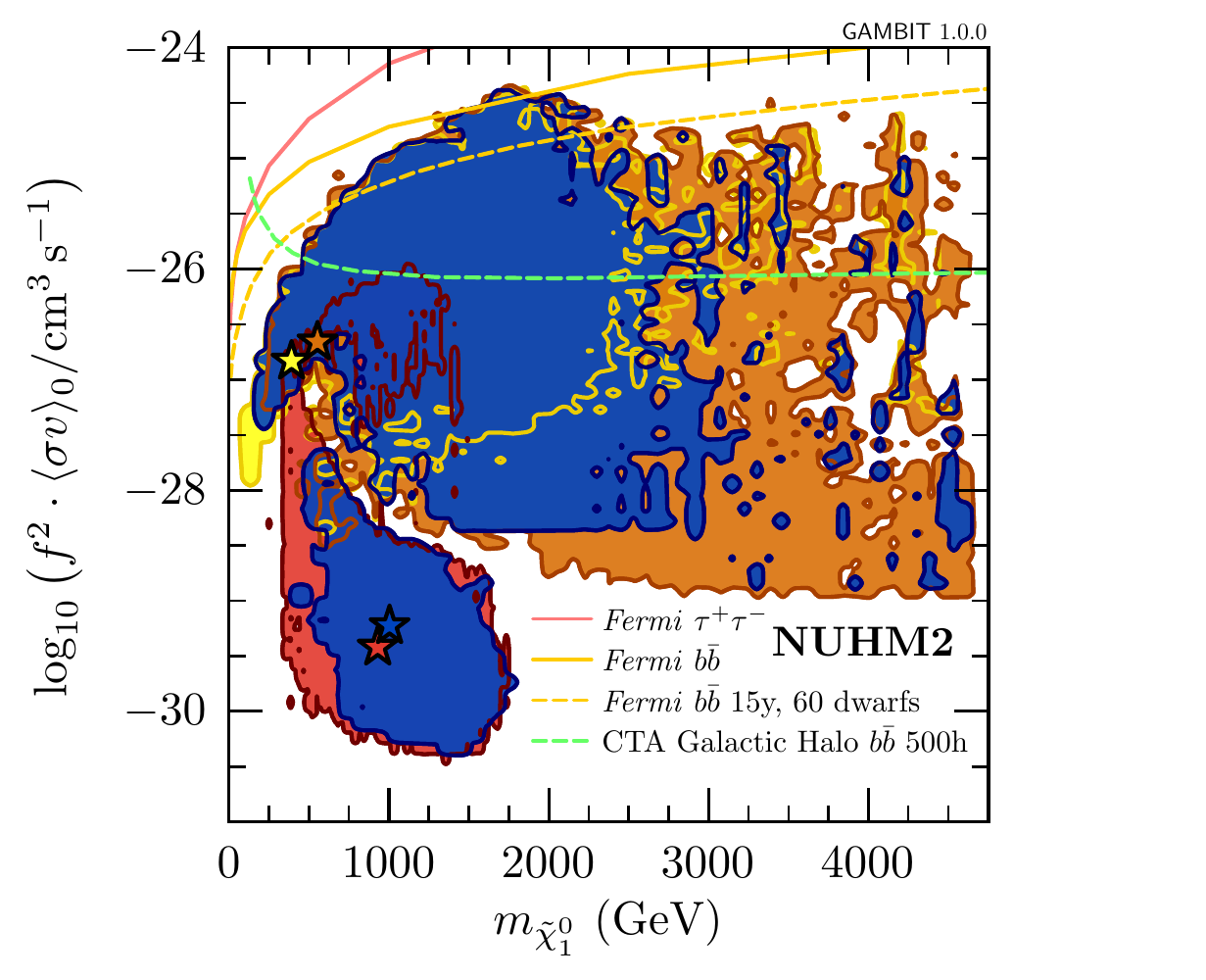}\\
\includegraphics[height=0.3cm]{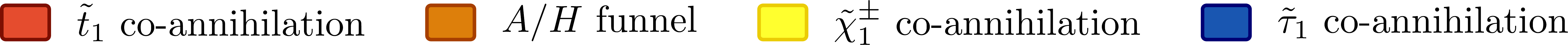}
\caption{As per Fig.\ \protect\ref{CMSSM}, but for the first and second Non-Universal Higgs Mass (NUHM) models.}
\label{NUHM}
\end{figure}

\begin{figure}
\includegraphics[width=0.49\textwidth]{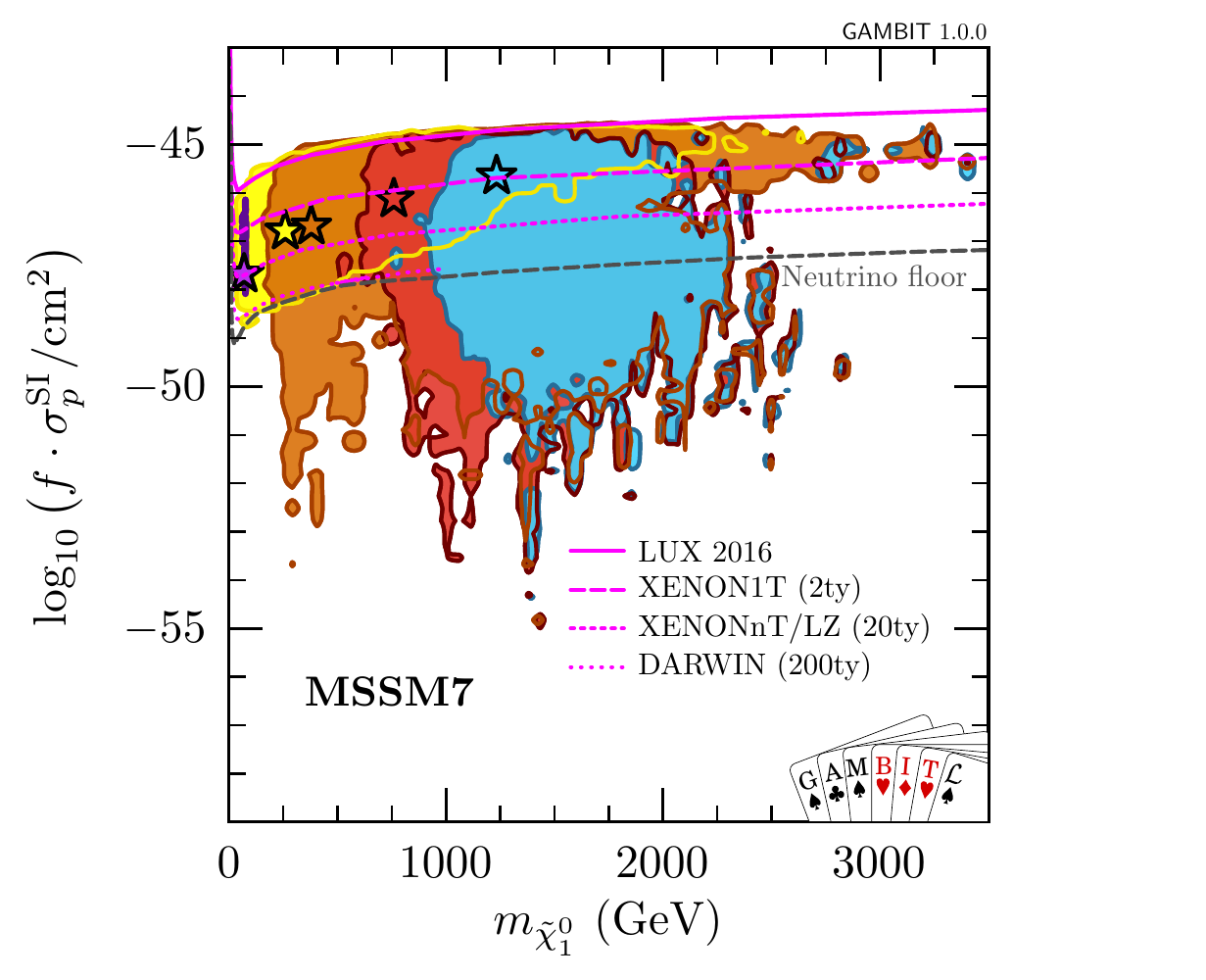}
\includegraphics[width=0.49\textwidth]{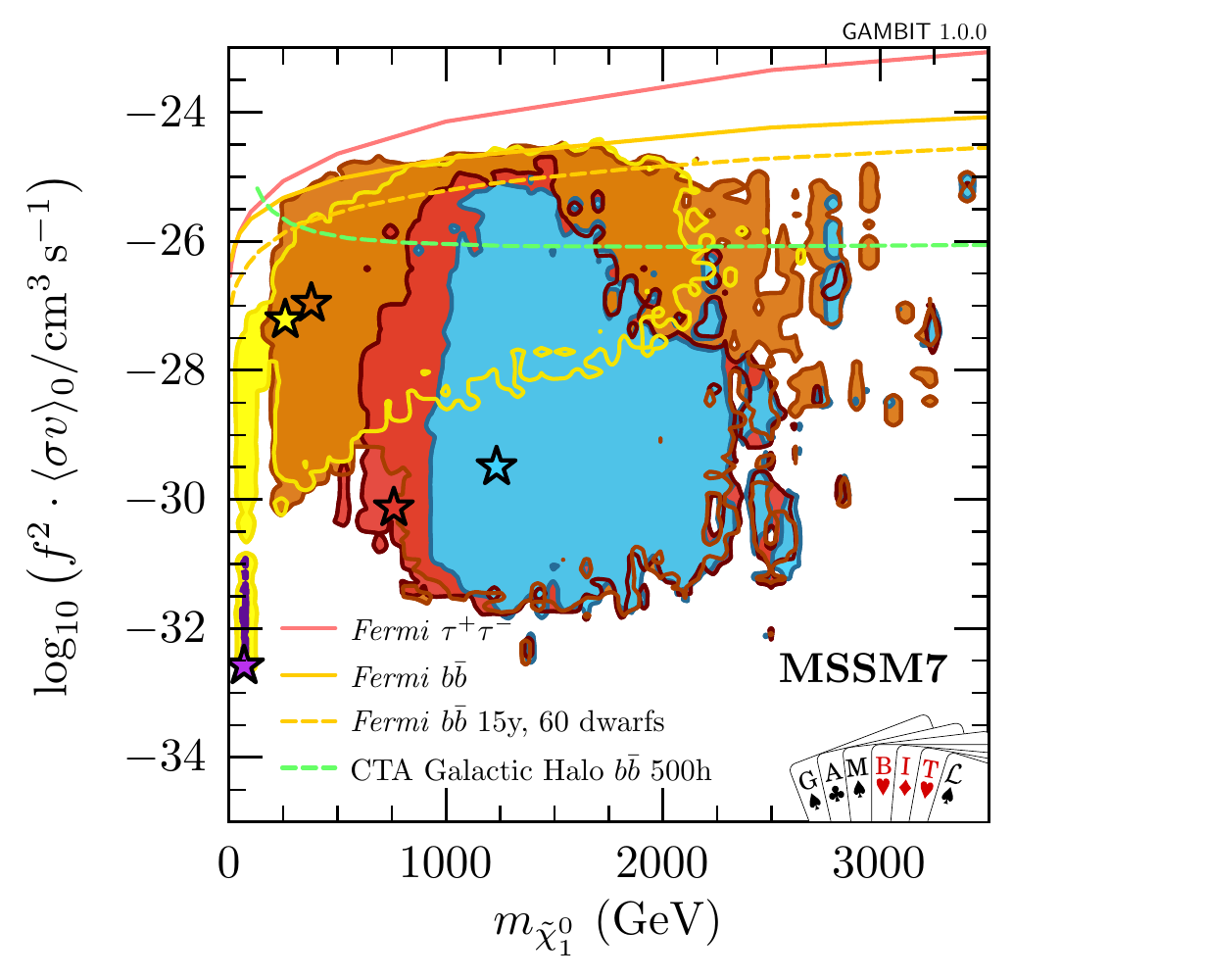}\\
\includegraphics[height=0.3cm]{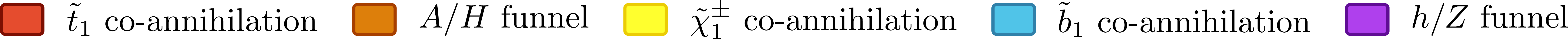}
\caption{As per Fig.\ \protect\ref{CMSSM}, but for the 7-parameter weak-scale phenomenological Minimal Supersymmetric SM (MSSM7) \cite{MSSM}.}
\label{MSSM7}
\end{figure}
\section{Supersymmetry and prospects for future discovery with \textit{Fermi} and CTA}

Similar analyses have been done in supersymmetry \cite{CMSSM,MSSM}.  Figs.\ \ref{CMSSM}--\ref{MSSM7} show 95\% confidence regions for the spin-independent nuclear scattering and annihilation cross-sections of DM in four supersymmetric models.  The cross-sections in Figs.\ \ref{CMSSM}--\ref{MSSM7} have all been rescaled on a per-model basis by $f$ (nuclear scattering) or $f^2$ (annihilation), to factor out the signal suppression caused by the relic density of the models.  This allows the preferred parameter regions to be consistently plotted alongside the respective experimental limits (which assume $f=1$).  The regions are colour-coded according to the specific mechanisms at play in depleting the relic density of DM, ranging from resonant annihilation via both light (SM) and heavy (MSSM) Higgs bosons, to various sfermion co-annihilation scenarios, and dominantly Higgsino DM (marked as `$\chi^\pm_1$ co-annihilation').

The first three models (\cite{CMSSM}; Figs.\ \ref{CMSSM} and \ref{NUHM}) are defined at the grand unification scale.  The NUHM2 (Non-Universal Higgs Mass 2) has one more parameter than the NUHM1, which has one more than the CMSSM (Constrained MSSM; Minimal Supersymmetric SM).  The final model (Fig.\ \ref{MSSM7}) is defined at the weak scale, in terms of 7 unified Lagrangian parameters (\cite{MSSM}; MSSM7).

Figs.\ \ref{CMSSM}--\ref{MSSM7} include the official limits from \textit{Fermi} searches for DM annihilation in dwarf spher-oidals \cite{LATdwarfP8}.  They clearly have no impact on the allowed parameter space of the CMSSM (Fig.\ \ref{CMSSM}).  As the model freedom is expanded in the NUHM1 and NUHM2 (Fig.\ \ref{NUHM}), a broader range of cross-sections is allowed.  In the weak-scale MSSM7 (Fig.\ \ref{MSSM7}), even more of the space remains viable.  The NUHM2 and MSSM7 have a number of models that annihilate via a heavy Higgs resonance, where the late-time cross-section is boosted relative to the effective value during freezeout.  This gives a present-day cross-section well above the canonical thermal value of $3\times10^{-26}$\,cm$^3$\,s$^{-1}$, but a relic density in agreement with the observed value. Figs.\ \ref{NUHM} (lower) and \ref{MSSM7} show that current searches with \textit{Fermi} already provide the most important constraints on many such models.

The righthand panels of Figs.\ \ref{CMSSM}--\ref{MSSM7} also show the projected sensitivity of \textit{Fermi} after 15 years observing 60 dwarf spheroidal galaxies \cite{Charles:2016pgz}.  Future \textit{Fermi} searches will probe many presently unconstrained models, still allowing good prospects for unambiguously detecting DM.  However, comparison with current constraints and future sensitivities of upcoming direct detection experiments, given in the left panels of these figures, shows that \textit{Fermi} will have strong competition from ton-scale liquid noble gas detectors, which will probe many of the same models in the near future.

The upcoming Cherenkov Telescope Array (CTA; \cite{Carr:2015hta}) will search for $\gamma$ rays from DM annihilation.  Figs.\ \ref{CMSSM}--\ref{MSSM7} show the projected sensitivity \cite{Carr:2015hta}.  Although these projections are rather optimistic \cite{Silverwood14,Pierre14}, particularly as they ignore the impacts of systematics expected to dominate the final limits \cite{Silverwood14}, CTA will probe a substantial amount of presently-viable supersymmetric model parameter space. Similar conclusions can also be drawn in effective field theories for DM \cite{2017PhRvD..96h3002B}.

\section{Summary}

To uncover the particle nature of DM, it is important to consider implications of specific DM models. To do this accurately requires per-model calculation many observables, theoretical implications and experimental limits, across the full parameter spaces of a number of different theories.

Scalar singlet DM is now constrained to two regions: a small `resonant triangle' at $m_S\lesssim m_h/2$, and a high-mass region presently beyond the reach of direct detection.  \textit{Fermi} searches for DM annihilation in dwarf spheriodal galaxies place leading constraints on this model specifically at the highest-mass edge of the resonant region.  Searches for DM annihilation in dwarfs with \textit{Fermi} also place significant constraints on supersymmetric models such as the NUHM and weak-scale MSSM.  Models where the relic density is depleted to the observed value by annihilation via a heavy Higgs resonance will be especially interesting for \textit{Fermi} and CTA, as many such models are presently unconstrained by other searches, and will be susceptible to discovery by one or both $\gamma$-ray experiments in the coming years.  Any detection of such models in $\gamma$ rays will however require quite rapid analysis by the respective experimental teams, and careful cross-checking against results from direct detection experiments, as LZ and XENON are expected to become sensitive to most such models on comparable or shorter timescales than either \textit{Fermi} or CTA.

\textbf{Acknowledgements}: I am supported by STFC (ST/K00414X/1, ST/P000762/1, ST/L00044X/1), and thank my co-authors on a number of the works discussed here.

\textbf{Questions.} \textit{Manuel Meyer}: Can GAMBIT characterise the true model if a signal is found in some experiment, rather than just limits?  \textit{Answer}: Yes, \GB was designed with this specifically in mind; see the effective field theory fit to current flavour anomalies \cite{FlavBit}, for example.

\textit{Manuel Meyer}: What drives the apparent limit on the parameter space from below in e.g. Figs.\ \ref{CMSSM}--\ref{MSSM7}, given that these regions correspond to small couplings, where naively I would expect no experiment to have sensitivity? \textit{Answer}: Different constraints are complementary, and do not all act from above when viewed in the planes of Figs.\ \ref{CMSSM}--\ref{MSSM7}.  For example, the LHC measurement of the Higgs mass is highly constraining, but it maps to the planes in Figs.\ \ref{CMSSM}--\ref{MSSM7} in a highly non-trivial way.  Similar for other collider, flavour and precision constraints.  The relic density also plays a significant role in restricting models to the regions seen in these figures, from multiple directions.

\textit{Alessandro Cuoco}: Do \textbf{all} the models you find in the resonance region of the singlet model have sub-dominant relic densities? \textit{Answer}: No, there are models in the resonance region, at the bottom of Fig.\ \ref{singlet}, where the singlet constitutes all the DM.

\bibliographystyle{JHEP_pat}
\bibliography{DMbiblio,SUSYbiblio}

\end{document}